\documentclass[12pt,tightenlines,eqsecnum,floats,showpacs,nofootinbib,amsmath,amssymb,aps,prd]{revtex4-2}
\usepackage[english]{babel}

\usepackage{amsmath}
\usepackage{hyperref}
\usepackage[usenames,dvipsnames]{xcolor}
\hypersetup{colorlinks=true, citecolor=Plum, linkcolor=Plum,
urlcolor=Plum}

\usepackage{tikz}
\usepackage{graphicx,verbatim}
\usepackage{amsfonts}
\usepackage{amssymb}
\usepackage{psfrag}
\usepackage{accents}
\usepackage{mathrsfs}
\usepackage{enumitem}
\usepackage{wasysym}

\def\be{\nopagebreak[3]\begin{equation}}
\def\ee{\end{equation}}
\def\ba{\nopagebreak[3]\begin{eqnarray}}
\def\ea{\end{eqnarray}}
\newcommand{\f}{\frac}
\def\rmd{\rm d}
\def\lp{\ell_{\rm Pl}}

\def\t{\tilde}
\def\h{\hat}
\def\uDelta{\underline{\Delta}}

\def\TDH {\emph{T-DH}\, }
\def\ATDH {\emph{AT-DH}\, }

\def\T{\mathcal{T}}

\newcommand*{\scri}{\ensuremath{\mathscr{I}}} 
\newcommand*{\scrip}{\ensuremath{\mathscr{I}^{+}}} 
\newcommand*{\scrim}{\ensuremath{\mathscr{I}^{-}}}

\begin{document}

\title{Black Hole Evaporation in Loop Quantum Gravity}
\author{Abhay Ashtekar}
\affiliation{Physics Department, Penn State, University Park, PA 16802, USA}
\affiliation{Perimeter Institute for Theoretical Physics, 31 Caroline St. N, Waterloo, ON N2L 2Y5, Canada}

\begin{abstract}

The conference \emph{Black Holes Inside and Out} marked the 50th anniversary of Hawking's seminal paper on black hole radiance. It was clear already from Hawking's analysis that a proper quantum gravity theory would be essential for a more complete understanding of the evaporation process. This task was undertaken in Loop Quantum Gravity (LQG) two decades ago and by now the literature on the subject is quite rich. The goal of this contribution is to summarize a mainstream perspective that has emerged. The intended audience is the broader gravitational physics community, rather than quantum gravity experts. Therefore, the emphasis is on conceptual issues, especially on the key features that distinguish the LQG approach, and on concrete results that underlie the paradigm that has emerged.  This is \emph{not} meant to be an exhaustive review. Rather, it is a broad-brush stroke portrait of the present status. Further details can be found in the references listed.
\end{abstract}

\maketitle

\section{Introduction}
\label{s1}
Hawking's discovery \cite{swh1} that isolated black holes emit quantum radiation which is approximately thermal at late times used  quantum field theory on a classical, background space-time of a collapsing star (see the left panel of Fig. 1). The calculation involved three key approximations: \emph{(i)} Space-time geometry can be treated classically; \emph{(ii)} Quantum fields can be regarded as `test fields' so that their back reaction on space-time geometry can be ignored; and, \emph{(iii)} The matter field which collapses is classical, \emph{distinct} from the test quantum field considered. 
\footnote{\label{fn1} The third assumption is not emphasized in the literature probably because it is harmless in the external field approximation. However, it is a rather severe limitation in the subsequent discussion of `information loss', or, of `unitarity of quantum dynamics'. For these considerations, we need a closed system: we need to able to specify the full incoming \emph{quantum} state --including that of collapsing matter-- on $\scrim$.}% (or on $\scrim \cup \inot$).}
A truly remarkable calculation then showed that, if the incoming state on $\scri^{-}$ for the quantum field is the vacuum, the outgoing state at $\scri^{+}$ is a mixed state which, at late times, is thermal. 
 \begin{figure} 
  \begin{center}
    \begin{minipage}{1.5in}
      \begin{center}
        \includegraphics[width=1.3in,height=2.2in,angle=0]{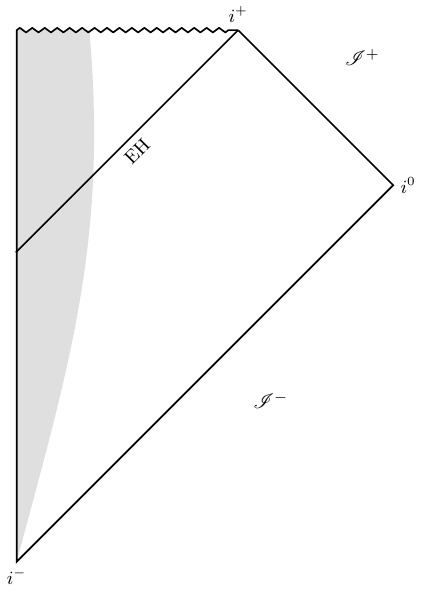} \\ {(a)}
         \end{center}
    \end{minipage}
   \hspace{.5in}
    \begin{minipage}{1.5in}
      \begin{center}
       \includegraphics[width=1.3in,height=2.4in,angle=0]{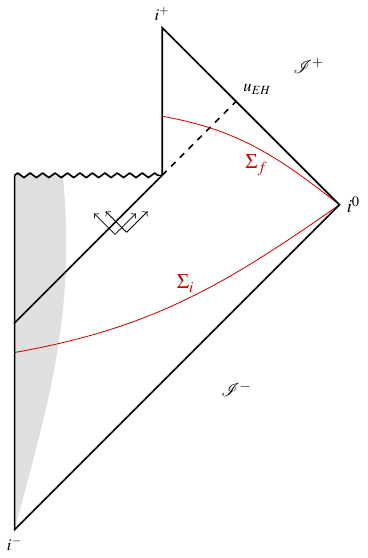} \\ {(b)}
        \end{center} 
   \end{minipage}
      \hspace{.5in}
       \begin{minipage}{1.5in}
        \begin{center}
         \includegraphics[width=1.3in,height=2.2in,angle=0]{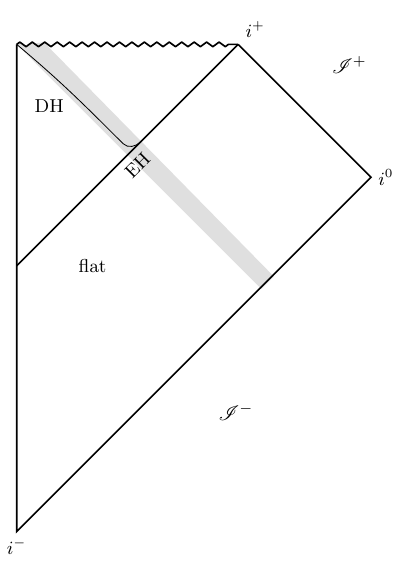} \\ {(c)}
          \end{center}
  \end{minipage}
\caption{\footnotesize{ (a) \emph{Left Panel}: The Penrose diagram of a collapsing star, used by Hawking in his calculation in the external potential approximation.
(b) \emph{Middle Panel}: The Penrose diagram proposed by Hawking in 1975 to incorporate the back reaction of an evaporation black hole.
(c) \emph{Right Panel}: The classical Vaidya space-time depicting a null-fluid collapse.}}
\vskip-1cm
\label{fig:1}
\end{center}
\end{figure}
%\vskip-0.5in
To remove assumption \emph{(ii)}, one has to include the back reaction of the quantum radiation on the classical space-time geometry. A natural set of heuristics led Hawking to propose that when the back reaction is included, the space-time diagram in the  Panel (a) of Fig. 1 would be replaced by the one in Panel (b): To compensate for the mass lost through quantum radiation, the black hole would lose mass and the singularity would not extend all the way to $\scrip$ but terminate in the space-time interior, leaving us with a Minkowski metric in the upper triangular region. In this depiction, a singularity persists and serves as a `sink of information'. Put differently, because of the singularity, it is clear that $\scrip$ would  not be the entire future boundary of space-time and therefore the S-matrix from $\scrim$ to $\scrip$ would not be unitary but would have to be replaced by the so-called \$ matrix \cite{swh2}. Even though 5 decades have passed since Hawking's discovery, we still do not have a detailed calculation of the back reaction even today. But the Panel (b) of Fig.1 continues to be widely used. Interestingly, this is done both to argue that information is lost \cite{uw} and to construct mechanisms for it not to be lost (but to appear on the portion of $\scrip$ to the past of the retarded time $u_{EH}$ in the figure) \cite{marolf,amps}! 

The LQG perspective is that this Penrose diagram is incorrect. In its place, a new paradigm was introduced about 2 decades ago \cite{aamb} and most of the LQG work on black hole evaporation has been devoted to develop it systematically through innovative constructions and detailed calculations. The paradigm is based on two key observations:\vskip0.05cm
\emph{(1)} The emphasis on event horizons (EHs) that the Penrose diagram in Panel (b) places is not justified in quantum gravity. To determine if a space-time admits an EH, one needs complete knowledge of space-time geometry of the evaporating black hole to infinite future and only full quantum gravity can provide this information. One should not \emph{presuppose} the nature of this geometry; one must let evolution equations of the appropriate theory determine it. As Matt Visser reminded us in his talk, in the GR-17 conference in Dublin, Hawking emphasized that ``a true event horizon never forms". This is precisely the LQG viewpoint: there is no event horizon either in the semi-classical space-time nor in the expected space-time geometry in full quantum gravity.\vskip0.05cm  
\emph{(2)} Singularities of classical general relativity are signposts of the limitation of classical general relativity, and gates to physics beyond Einstein. In LQG there are strong reasons to expect that all space-like, strong curvature singularities will be naturally resolved by quantum geometry effects. Consequently, the quantum space-time would extend beyond the putative classical singularity, providing a path way for information recovery on the larger $\scrip$  of the quantum extended space-time.\\
It is interesting to note that Hawking changed his mind about the Panel (b) of Fig.1 in 2016  and replaced it with a Penrose diagram that has neither a singularity nor an event horizon \cite{hps}!

This article is organized as follows. In section \ref{s2} we discuss in some detail main results that motivate the LQG perspective on black hole horizons and singularities. Specifically, we will address the following questions:  (i) What should replace EHs in the description of black holes in classical and quantum gravity? What is it that forms in a gravitational collapse and evaporates due to quantum emission? and, (ii) How does quantum Riemannian geometry of LQG lead to the resolution of the black hole singularity? What is the nature of the geometry in the quantum extension of space-time in the Planck regime? In section \ref{s3}, we will use these concrete results as pointers to suggest a pathway to address apparent paradoxes that arise in the analysis of the issue of information loss, first in the semi-classical theory and then in full quantum gravity. In section 4 we will summarize the main results and discuss the issues that remain. 

There is inevitable overlap with review articles \cite{aa-eva,aosrev,aaeb,akrev2}, where further details can be found.

\section{The new elements of the LQG Perspective} 
\label{s2}

As explained in section \ref{s1}, there are two key points on which LQG investigations differ from the  commonly used narrative enshrined in Hawking's original paradigm depicted in the Panel (b) of Fig. \ref{fig:1}. In this section we provide further details. In the first part, \ref{s2.1}, we discuss quasi-local horizons  that can be used to represent black holes both in the formation and evaporation process. In the second part, \ref{s2.2}, we discuss the resolution of the Schwarzschild singularity through quantum geometry effects of LQG and summarize the key features of the resulting quantum corrected geometry.

\subsection{From EHs to Quasi-local Horizons}
\label{s2.1}
In the 1970s when black holes were first analyzed using global techniques, EHs played a seminal role and led to several interesting results. Perhaps the most important among them is the area law established by Hawking \cite{swh3} that made the similarity between the laws of black hole mechanics and laws of thermodynamics compelling. Therefore subsequently, black holes were generally regarded as synonymous with EHs. 

However, as we already noted, the notion is very global: EH is the future boundary of the space-time region $J^-(\scrip)$ from which future directed causal signals can be sent to $\scrip$. For the notion to be non-trivial, $\scrip$ has to be complete in a standard sense \cite{rggh} (since if we allow incomplete $\scrip$, even Minkowski space could be said to have an EH!). Therefore to determine if a space-time admits an EH, and then to locate it, one needs the knowledge of space-time geometry to the infinite future. Already in classical general relativity, this feature prevents one from using  EHs \emph{during} numerical simulations of binary black hole mergers: One cannot use EHs to locate the progenitor black holes, nor to say when the merger happens, nor to locate and study properties of the final black hole remnant. In these simulations the EH can only be reconstructed only \emph{at the end}, as an afterthought thought. In quantum gravity, the role of EHs is even more dicey: since as of now we do not have even the adequate equations to construct the complete space-time of an evaporating black hole, we cannot look for it even as an afterthought! 

It has long since been realized that EHs have an even more severe limitation: they are teleological and therefore inherit some spooky features. For example, an EH may well have just formed and grown in size in the very room where you are sitting, in \emph{anticipation} of a gravitational collapse in this region of our galaxy a billion years from now.  This feature is concretely realized in the null fluid collapse of a Vaidya space-time shown in the  Panel (c) of  Fig. 1, where the event horizon forms and grows in the flat region of space-time, in anticipation of the null fluid collapse, even though nothing at all is happening in the flat part of space-time. And null fluids are not necessary to illustrate this limitation of EHs. Recent results of Kahle and Ungar show that the same phenomenon occurs if one uses Vlasov fluids (that emerge from $i^-$ rather than $\scrim$, see Fig. 1 of \cite{kh}). 

These limitation motivated the introduction of \emph{quasi-local horizons} (QLHs) to describe black holes and their dynamics in a manner that not only avoids teleology but leads to a framework that is directly useful in simulations to locate black holes and to extract physics from numerical outputs. (See in particular \cite{hayward1,ih-prl,ak1,ak2,akrev,boothrev,jjrev}.) In the black hole evaporation process, one can locate the QLHs knowing only the part of space-time where semi-classical approximation holds (i.e., the the black hole continues to be macroscopic) and relate the changes in their geometry to physical processes. One does not need to know the space-time to infinite future. And, as we will see, some qualitatively new features arise already at the semi-classical level. 

\subsubsection{QLHs in classical general relativity}
\label{s2.1.1}

One begins with the notion of a \emph{marginally trapped surface}(MTs) $S$: a closed 2-manifold (which we will take to be topologically $\mathbb{S}^2$) for which one of the (future-directed) null normals, say $\ell^a$, is expansion-free: $\theta_{(\ell)} =0$. The notion is quasi-local rather than local because the expansion has to vanish on the entire 2-sphere $S$. A QLH is a 3-manifold $\mathfrak{H}$ that is foliated by MTTs. (Therefore QLHs have also been referred to as marginally trapped tubes (MTTs) \cite{aagg}.) In striking contrast to EHs, the definition of QLHs refers only to the space-time geometry in an infinitesimal neighborhood of $\mathfrak{H}$ and all their properties can be deduced just from this local geometry. Therefore there is no teleology. In particular, you can rest assured that there is no QLH contained in your room!

Of particular interest to the discussion of black hole evaporation is a sub-case of QLHs called \emph{dynamical horizons} (DHs) $H$ that have the following properties: (i) they are nowhere null;  (ii) the expansion $\theta_{(n)}$ of the other null normal $n^a$ to the MTSs is nowhere vanishing; and (iii) $\mathcal{E} := |\sigma|^2 + 8\pi G\, T_{ab} \ell^a \ell^b$ does not vanish identically on $H$ \cite{akrev2}. One can regard $\mathcal{E}$ as a certain type of energy flux. In the panel (c) of Fig.~\ref{fig:1}, for example, a (space-like) DH develops during there collapse when $\mathcal{E} >0$. At the end of the infall, $\mathcal{E}$ vanishes and the QHL --depicted by the null segment-- becomes non-dynamical. 
\footnote{This notion differs from the original one \cite{ak1,ak2} which was geared to the study of QLHs that represent remnants in the numerical simulations of black hole mergers. The one used in this review accommodates evaporating black holes as well. It can be further generalized by weakening the `nowhere null' condition. We use the stronger notion just to keep the discussion brief.}

\begin{figure} 
  \begin{center}
     \begin{minipage}{1.5in}
      \begin{center}
       \includegraphics[width=1.9in,height=2.1in]{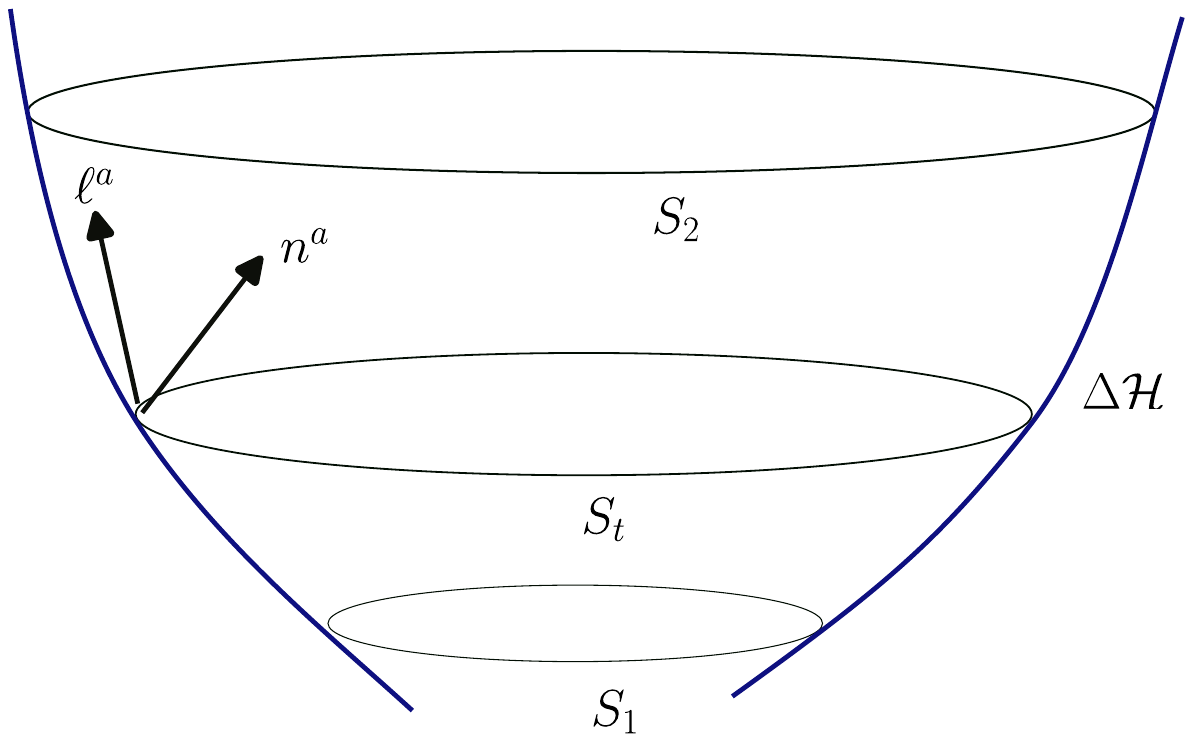}\\ (a)
         \end{center}
    \end{minipage}
   \hspace{.5in}
    \begin{minipage}{1.7in}
      \begin{center}
       \includegraphics[width=1.7in,height=1.7in,angle=0]{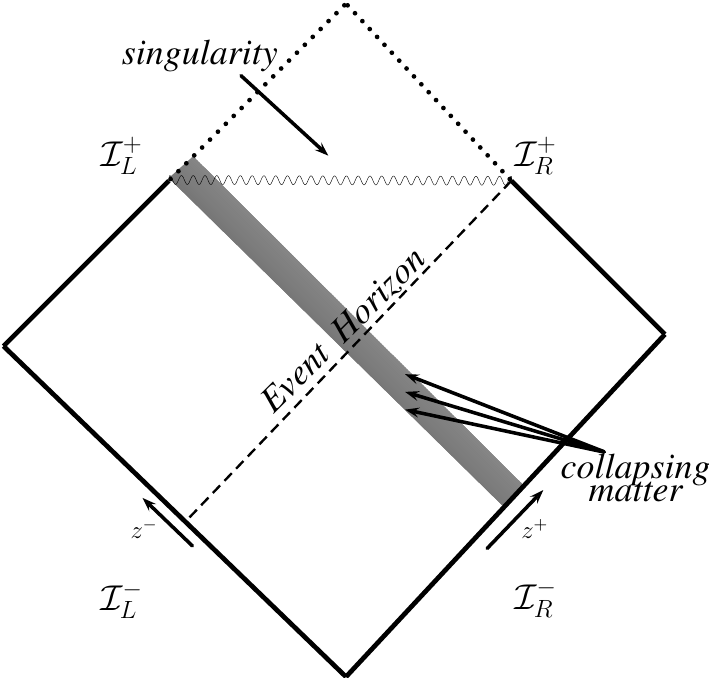} \\ {(b)}
        \end{center} 
   \end{minipage}
      \hspace{.5in}
       \begin{minipage}{1.7in}
        \begin{center}
         \includegraphics[width=1.7in,height=1.7in,angle=0]{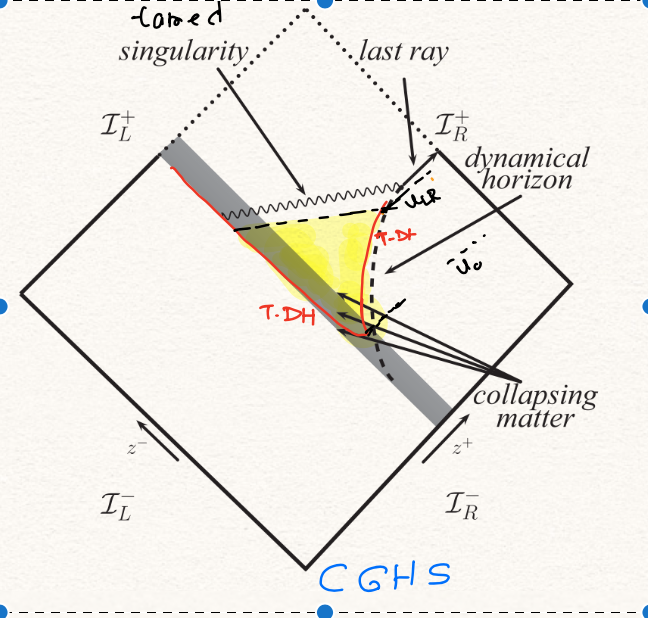} \\ {(c)}
          \end{center}
  \end{minipage}
\caption{\footnotesize{ (a) \emph{Left Panel}: A dynamical horizon $H$, foliated by marginally trapped surfaces. A typical leaves are marked $S_t$ and two null normals to it are denoted by $\ell^a$ and $n^a$. $\Delta H$ is the portion of the DH bounded any $S_1$ and $S_2$. (b) \emph{Middle Panel}: The Penrose diagram of a classical 2-d black hole formed by the collapse of a massless scalar field coming in from $\scrim$ depicted by the shaded region. Space-time metric is flat to the past of the shaded region and represents a static black hole to the future. Again, the event horizon forms and evolves in the flat region of space-time. (c) \emph{Right Panel}: The semi-classical extension of the space-time in panel (b). During the collapse, a trapping dynamical horizon T-DH (depicted in red) forms, is space-like, and grows as the matter collapses. At the end of this process, T-DH becomes time-like and starts shrinking in area due to the influx of negative energy. The shaded yellow region is semi-classical and trapped. To its future one needs full quantum gravity. }}
\vskip-0.5cm
\label{fig:2}
\end{center}
\end{figure}

It follows immediately from the definition of a DH that the area of MTSs is a monotonic function on $H$. Therefore we can use the areal radius $R$ as a coordinate on $H$ so that the MTSs are given by $R= {\rm const}$. Interestingly, one can now show that the change in area is directly related to the physical processes \emph{occurring at the DH}. For example, in the case when the DH is space-like, as in the Panel (a) of Fig.~\ref{fig:2}, area increases along the projection of $\ell^a$ into the DH and the difference in the areal radius $R_1$ and $R_2$ of two MTSs is given by the `flux of energy' into the portion $\Delta H$ of the DH bounded by the two cross-sections \cite{ak1,ak2}:
\be \label{balance1} \f{1}{2G}\, (R_2 - R_1) =  \underbrace{\int_{\Delta H} N\, T_{ab} \ell^a \hat{\tau}^b \, \rmd^3V}_{\hbox{{\rm Matter energy flux}}} + \underbrace{\f{1}{16\pi G}\, \int_{\Delta H} N\, \left(
|\sigma|^2 + 2 |\zeta|^2 \right)\, \rmd^3V}_{\hbox{\rm GW\, energy flux}} \ee 
\vskip0.1cm
\noindent Here `energy' is defined relative to the causal vector field $N\ell^a$, where $N = |\partial_R|$ (recall: $\ell^a$ is the distinguished causal vector field on $H$). There is a similar formula in the case when $H$ is time-like, again only involving physical processes on the portion $\Delta H$ of $H$,  but now the area \emph{decreases} along the projection of $\ell^a$ into $H$ \cite{ak2}. Thus, not only does the key `area law' that elevated the status of EHs in the 1970s continue to hold for DHs, \emph{but it holds in a stronger sense}. For EHs we only have a qualitative statement that the area cannot decrease. As the example of Vaidya space-time vividly brings out, the change need not be related to any physical processes near the EH since the area can grow even in flat space-time. This does not happen with DHs: Now we have a \emph{quantitative} statement (\ref{balance1}) that relates the change in area to local \emph{physical processes} occurring on the relevant portion $\Delta H$ of $H$. In particular, there are no DHs whose 2-sphere cross-sections lie in flat space-time, in striking contrast to EHs.
\smallskip

\emph{Remark:} Note that DHs  are 3-dimensional sub-manifolds $H$ in space-time in their own right, 
defined using structures that refer just to these sub-manifolds; in particular, one does not need a 3+1 slicing of space-time by Cauchy surfaces. The foliation by MTSs is intrinsic to $H$. Furthermore, the foliation on any given DH is unique --$H$ does not admit two distinct foliations by MTSs \cite{aagg}.  Thus the notion of a DH is distinct from that of an \emph{apparent horizon} --the `outermost' MTS on a Cauchy surface. Unfortunately, the term `apparent horizon' is often employed in the contemporary literature to emphasize that the structure used is different from an EH or a Killing horizon, when what one actually has is either an MTS or a DH, without any reference to a Cauchy surface! While the authors are well aware of which structures are actually used, this misuse of terminology can cause unnecessary confusion. Indeed, this occurred in the BHIO conference in 2 talks \cite{wald,visser}. 
\smallskip

For completeness, let us consider the complementary QLHs $\mathfrak{H}$ where the underlying 
{\smash{3-manifold}} is null. These QLHs are called  \emph{non-expanding horizons} (NEHs)  and denoted by $\Delta$. In this case, the null normal $\ell^a$ is also tangential to $\Delta$; \emph{every} cross-section of $\Delta$ is expansion-free. Furthermore the intrinsic (degenerate) metric on $\Delta$ is Lie-dragged by $\ell^a$ if null energy condition holds or we have spherical symmetry. %Of particular interest is the sub-case of NEHs is \emph{isolated horizons} on which $\ell^a$ also Lie-drags the intrinsic derivative operator on $\Delta$ \cite{ih-prl}. As the name suggests, in the context of black holes, one can think of IHs as depicting boundaries of black holes that are in equilibrium. They are more general than Killing horizons. For example, the Robinson-Trautman and Kastor-Traschen solutions admit IHs although there is no Killing field with respect to which they can be Killing horizons \cite{pc,kt}. 
Only the DHs that are directly relevant to the analysis of the black hole evaporation process because there are no fluxes of energy $\mathcal{E}$ across an NEH. But in the long semi-classical phase of black hole evaporation where the flux into the horizon is very small,  the exact DH is well-approximated by a perturbed NEH \cite{akkl2}.

\subsubsection{DHs in semi-classical gravity: An example}
\label{s2.1.2}

As remarked in footnote \ref{fn1}, a proper analysis of the issue of `information loss' and `unitarity of the S-matrix', requires a closed system. The widely considered case of a stellar collapse does not  satisfy this criterion because it is very difficult, if not impossible, to specify the incoming \emph{quantum} state of the star in the analysis. However, this can be rectified by considering a massless scalar field collapse and using, say a coherent state peaked at an appropriate infalling classical scalar field that collapses to form a black hole. Therefore from now on we will focus on this case. 

Now, already in classical general relativity, the problem of studying this collapse is quite complicated especially because of the associated critical phenomena. But there are some slightly simplified models that are in fact exactly soluble. Perhaps the most well-known among these is the Callen, Giddings, Harvey, Strominger (CGHS) model \cite{cghs}. It describes the spherical collapse of a scalar field in a 4-d theory of gravity that differs from general relativity. However, as is well-known, one can reduce the spherical symmetric sectors of such 4-d theories to  2-d theories in which the dynamical variables are the 2-d metric $g_{ab}$, a dilation $\phi$ (where $R = e^{-\phi}$ is the radius of 2-spheres in the 4-d theory), and massless scalar fields $f$; the first two capturing the information in the 4-d metric, and the third denoting the 4-d scalar field. In the spherically symmetric sector, the CGHS action differs from that of the spherical reduction of general relativity in a seemingly minor way --two of the terms have different coefficients. But technically these differences make a huge difference: in the classical theory, calculations trivialize in the CGHS model because one can decouple dynamics of the scalar field $f$ from that of geometric fields, the metric $g_{ab}$ and the dilation $\phi$. 

Further simplifications occur because the 2-d metric $g_{ab}$ is conformally flat and the 2-d massless Klein Gordon equation is conformally invariant. Therefore one can first trivially solve the wave equation for $f$, say for a left moving wave packet in 2-dimensional Minkowski space $(\mathring{M}, \mathring{g}_{ab})$ (shown as a diamond, including the dashed lines in the Panel (b) of Fig. \ref{fig:2}). Given this solution $f$, one can write down expressions of the dilation field $\phi$ and a metric $g_{ab}$, both constructed using certain integrals involving $f$. One can then verify that $f$ solves the wave equation also w.r.t. $g_{ab}$, and $(g_{ab}, \phi)$ satisfy the correct field equations with $f$ as a scalar field source. However, the metric $g_{ab}$ is now such that it has a space-like singularity (shown in the Panel (b) of Fig. \ref{fig:2}). Therefore, the physical space-time $(M, g_{ab})$ is only a portion of the Minkowski space $(\mathring{M}, \mathring{g}_{ab})$ we began with (that lies to the past of the wiggly line depicting the singularity)! In particular, then, the $\scrip$ of $g_{ab}$ is a \emph{proper subset} of the $\mathring{\scrip}$ of the underlying Minkowski space $(\mathring{M}, \mathring{g}_{ab})$ (depicted by solid lines). Yet, $\scrip$ is complete in the geometry endowed on $M$ by the physical metric $g_{ab}$, just as $\mathring{\scrip}$ is complete w.r.t. the Minkowski metric $\mathring{g}_{ab}$! Therefore one can examine the causal past $J^-(\scrip)$ to check if the physical space-time $(M, g_{ab})$ admits an event horizon. As the figure brings out, it does; the singularity is hidden behind this horizon. Thus, the solution $(g_{ab}, \phi, f)$ we have constructed, starting just with a solution to the wave equation in 2-d Minkowski space,  represents a black hole in this 2-d theory! (For a pedagogical self-contained summary, see e.g. section I.A of \cite{apr}.)

Using the facts that the model is exactly soluble in the classical theory, and we are in 2 space-time dimensions, it is possible to write down the semi-classical equations as well. The key approximation in the semi-classical theory is that the quantum metric operator can be replaced by its expectation value. Thus in this theory one has a smooth, `c-number'  metric (but with coefficients that depend on $\hbar$) sourced by the quantum matter. This approximation can be justified if one has a large number $N$ of scalar fields in the problem, so that the quantum fluctuations in the geometric fields (the metric and the dilation) can be neglected compared to the total quantum fluctuations in $N$ scalar fields in a $1/N$ expansion \cite{cghs}. 

In this approximation, one has just a set of partial differential equations for the metric coefficients. However they are non-linear and rather complicated for an analytical treatment (although in certain regimes it is possible to obtain approximate analytical expressions \cite{ori2}). There is a long history of using numerical methods to solve them but, as we will discuss below, they led to some incorrect results because (i) the numerical precision was not sufficiently high; (ii) the treatment of a key conceptual issue (the definition of Bondi energy in the semi-classical theory) was flawed; and, (iii) an important scaling symmetry of the fundamental equations was not recognized. These limitations were addressed in \cite{aprlett,apr, ori2} and the careful analysis led to several qualitatively new results that can guide us in the investigation of black hole evaporation in 4 dimensions. These results can be summarized as follows:
%\vskip0.05cm

\emph{(1)} Simulations assumed that the quantum state of the infalling scalar field is a coherent state on $\scrim$, peaked at as classical scalar field $f$ of finite duration, that undergoes a prompt collapse. As the scalar field falls in, a trapping dynamical horizon is formed. It is space-like and grows in area in the support of the scalar field in space-time, in response to the infalling energy flux. Immediately after the infall ends, the DH turns around and becomes time-like and area of its MTSs starts shrinking due to the influx of negative energy across it that balances the outgoing positive energy flux across $\scrip$. Thus, \emph{what forms in the gravitational collapse and shrinks in the evaporation process is a DH}. (In this analysis, `area' and `expansions of null normals' used to identify the DH refer to the 4-d space-time from which the 2-d space-time is obtained using spherical symmetry reduction. In the 2-d picture their expressions involve the metric $g_{ab}$ as well as the dilation $\phi$ which encodes the radius of 2-spheres in the 4-d theory via $R= e^{-\phi}$.) 
\vskip0.05cm
\emph{(2)} \emph{There is no event horizon} in the semi-classical space-time. Numerics show that $\scrip$ of the semi-classical space-time is \emph{incomplete}. Therefore, as commented in section \ref{s2.1}, one cannot use the future boundary of $J^-(\scrip)$ as the event horizon. This future boundary represents `the last ray' labeled by $u_{LR}$ in panel (c) of Fig.~\ref{fig:2}.
\vskip0.05cm
\emph{(3)} The two branches of the DH, shown in red in the Panel (c) of Fig. \ref{fig:2}, enclose a trapped region, shown in yellow. The outer boundary of this region is time-like. Therefore one may be tempted to say that `information can leak out' across it and purify the quantum state at $\scrip$ before the last ray (as in \cite{hayward-conf}). However this possibility is not realized. There is no \emph{outward} flux across the time-like piece of the DH. Very recently, much more general analysis was carried out to argue that this possibility cannot be realized also in the spherical sector of general relativity (without having to use the CGHS-type simplifications) \cite{agulloetal}.
\vskip0.05cm
\emph{(4)} Both, analytical approximations \cite{ori2} and high precision numerics \cite{apr} show that the singularity is softened. The quantum corrected, semi-classical metric is continuous everywhere. However, one cannot trust the validity of the semi-classical approximation (i.e. the truncation of the $1/N$  expansion to leading order) once the curvature becomes Planckian. That is why the yellow semi-classical region in the figure does not extend all the way to the singularity.
\vskip0.05cm
\emph{(5)} While the singularity extends all the way to $\scrip$ in the classical theory (see Panel (b)  in Fig. \ref{fig:2}), it ends well inside the physical space-time in the semi-classical theory (see Panel (c) in Fig. \ref{fig:2}). There is a last ray that goes from the end of the singularity to $\scrip$ (marked $u_{LR}$ in Panel (c)). Hawking and Stewart had also analyzed this semi-classical space-time  numerically and concluded that there is a `thunderbolt singularity' along this last ray \cite{swhjs}. However, this conclusion appears to be a result of a simulations that did not have adequate accuracy. The high precision simulations carried out in \cite{apr} show that the metric is regular along this last ray; \emph{there is no thunderbolt}. There is also no trace of a `firewall' that was conjectured from string theory considerations \cite{amps}. Finally, using some plausible assumptions, it has been argued that the S-matrix would be unitary in full quantum theory \cite{atv}.
\vskip0.05cm
\emph{(6)} The retarded time at which the space-like branch of the DH ends and meets the time-like branch marks the onset of quantum radiation at $\scrip$. Furthermore, once the correct notion of Bondi-energy at $\scrip$ is identified for the semi-classical theory \cite{atv}, there is a direct correlation between the (shrinking) area of any given MTS on the time-like branch of the DH and the Bondi energy at the corresponding retarded time at $\scrip$. This `balance law' reinforces the physical significance of the DH. \vskip0.05cm

On the whole, then, these investigations have sharpened our expectations on what to expect in semi-classical limit and strongly indicate that suggestions of possible failures of semi-classical gravity in rather tame situations  are not viable. (The more recent findings of LIGO-Virgo collaboration have by now weeded out the suggestion that there would be of `firewalls' along horizons of macroscopic black holes.) There are also several other striking results obtained in the CGHS model (such as a scaling symmetry and universality of the Bondi mass at the end of the semi-classical evaporation process) that may be useful to the analysis of the evaporation in the spherically symmetric sector of general relativity. 

However, the CGHS model suffers from a fundamental limitation vis a vis this more realistic system: It is a genuinely 2-d model. First, the left and right $\scri^{\pm}$ are distinct. The collapsing scalar field originates on right $\scrim$ and moves left while the quantum radiation appears on right $\scrip$, originating in the vacuum state of right moving quantum fields on left $\scrim$. Second, in this 2-d model the temperature associated with the quantum radiation on right $\scrip$ is a constant, independent of the mass of the black hole (although there is an approach to improve on this situation using approximation methods \cite{ori3}). 

To rectify this situation, very recently Varadarajan has proposed a new model that is free from these limitations but is still manageable because the scalar field sector again decouples from gravity \cite{mv}. There is a single $\scrim$ and a single $\scrip$ and a symmetry axis on which the rotational Killing fields all vanish. The classical Penrose diagram is the same as in Panel (c) of Fig. \ref{fig:1}, rather than the Panel (b) of Fig. \ref{fig:2}. What is neglected vis a via spherical symmetric reduction of the collapse of a massless scalar field $f$ in general relativity is the \emph{back scattering} of $f$ by curvature which is responsible for the `gray body factors'. While this is a limitation, the model can be solved exactly in the classical theory and one can write down manageable semi-classical equations. So far, ignoring back scattering does not appear to be a severe drawback; it may just amount to the approximation in which the `gray body factors' are neglected.
Numerical simulations (that will be soon undertaken) should shed considerable light on this issue and also provide more reliable ways to test current scenarios of quantum evaporation.

\subsection{Singularity resolution in LQG}
\label{s2.2}

Let us now turn to the second issue on which LQG approaches depart decisively from the scenario advocated  in Hawking's original proposal, depicted in Panel (b) of Fig.\ref{fig:1}: fate of the Schwarzschild singularity in quantum gravity. As we discussed in section \ref{s1}, if the singularity persists as a part of the future boundary, information will be lost, i.e. the evolution of the quantum state from $\scrim$ to $\scrip$ would not be unitary. However, LQG results to date strongly suggest that this singularity would be resolved because of quantum geometry effects that lie at its foundations.

Let us  make a detour to describe of the nature of quantum geometry in LQG. To begin with, let us recall the central idea behind general relativity: Gravity is not a force as in Newton's theory but a manifestation of curvature of space-time. Therefore, to develop general relativity, Einstein had to use a new syntax to describe all of classical physics: differential geometry. Thus, space-time geometry is described by a metric, its derivatives operator and curvature, matter is represented by tensor fields that obey hyperbolic differential equations with respect to the metric. The LQG viewpoint is that one now needs a new syntax to formulate quantum gravity. Since gravity is encoded in geometry, a quantum theory of gravity should also be a quantum theory of geometry. Therefore, the new syntax is to be provided by \emph{quantum Riemannian geometry} in which basic geometric observables like areas of physical surfaces, volumes of physical regions and curvature of space-time are all represented by suitable operators. 

This syntax was created by a large number of researchers in the 1990s (for reviews, see, e.g., \cite{alrev,crbook,ttbook,kgrev}). The syntax is based on two key ideas: \emph{(i)} A reformulation of general relativity  (with matter) in the language of gauge theories  --that successfully describe the other three basis forces of Nature--  but now \emph{without reference to any background field}, not even a spacetime metric; and, \emph{(ii)} Subsequent passage to quantum theory using non-perturbative techniques from gauge theories --such as holonomies of the gravitational connection--  again without reference to  background fields. Consequently, the emphasis is shifted from metrics to connections. Background independence implies diffeomorphism covariance, which was heavily used together with non-perturbative methods \cite{lost}. One was then naturally led to a fundamental, in-built discreteness in geometry that foreshadows ultraviolet finiteness. The familiar spacetime continuum of general relativity is \emph{emergent} in two senses. First, it is built out of certain fields that feature naturally in gauge theories, without any reference to  a spacetime metric. Second,  it emerges only \emph{on coarse graining} of the fundamental discrete structures --the `atoms of geometry'-- of the quantum Riemannian framework. (For a recent short overview addressed to non-experts, see \cite{aaeb}.)

At a fundamental level, observables such as areas of surfaces and volumes of regions have a purely discrete spectrum. Thus geometry is `quantized' in the same sense that energy and angular momentum of the hydrogen atom are quantized. In particular, there is the smallest non-zero eigenvalue of the area operator, called the \emph{area gap} and denoted by ${\uDelta}$ whose value turns out to be $\sim  5.17$  in Planck units. It plays a key role in quantum dynamics. For, the curvature operator is defined by  considering the holonomy of the gravitational connection around a closed loop, dividing by the physical area enclosed by the loop, and then shrinking the loop until it encloses area ${\uDelta}$. Consequently, the curvature operator inherits a Planck scale non-locality from ${\uDelta}$ which in turn provides a natural ultra-violet regulator in quantum dynamics.

While full LQG is still being developed, its cosmological sector --\emph{Loop Quantum Cosmology} (LQC)-- has been investigated in great detail using non-perturbative methods of Hamiltonian LQG, the corresponding path integrals, as well as the consistent histories approach (for reviews, see, e.g., \cite{asrev,iapsrev}). These investigations have shown that the Big Bang Bang and the Big Crunch singularities of the homogeneous cosmologies  are naturally resolved by the quantum geometry effects. A key feature of the LQC dynamics is that corrections to general relativity are negligible until the matter density or curvature are $\sim 10^{-4}$ in Planck units, but then they grow very rapidly, creating an `effective repulsion' that completely overwhelm the classical attraction and causes the universe to bounce. In this singularity resolution, matter does not violate the standard energy conditions. Yet the singularity theorems in classical GR are bypassed because the quantum corrections modify Einstein's equations themselves. 

Many of the consequences of the LQC dynamics can be readily understood using the so-called \emph{effective equations} that capture the evolution of the peaks of sharply peaked quantum states (which, interestingly, remain sharply peaked also in the Planck regime). They encapsulate the leading order corrections to the classical Einstein's equation everywhere, \emph{including the Planck regime}. There is a streamlined procedure to arrive at the effective equations starting from full quantum dynamics that governs the quantum states in LQC \cite{vt,asrev}. The full dynamics of quantum states has much more detailed information. Effective equations extract from quantum states smooth metrics with coefficients that depend on $\hbar$, capturing the most important quantum corrections to dynamics. Therefore  they have been  heavily used to gain  valuable physical insights into the nature of quantum corrected geometry in the Planck regime. Note that the term `effective' is used in LQC in a sense that is \emph{very different} from the common usage in quantum field theory. In particular, there is no `integration of the ultraviolet degrees of freedom'; the LQC effective equations hold even in the Planck regime.
 \begin{figure} 
 
  \begin{center}
    \begin{minipage}{1.5in}
      \begin{center}
        \includegraphics[width=1.9in,height=2.2in,angle=270]{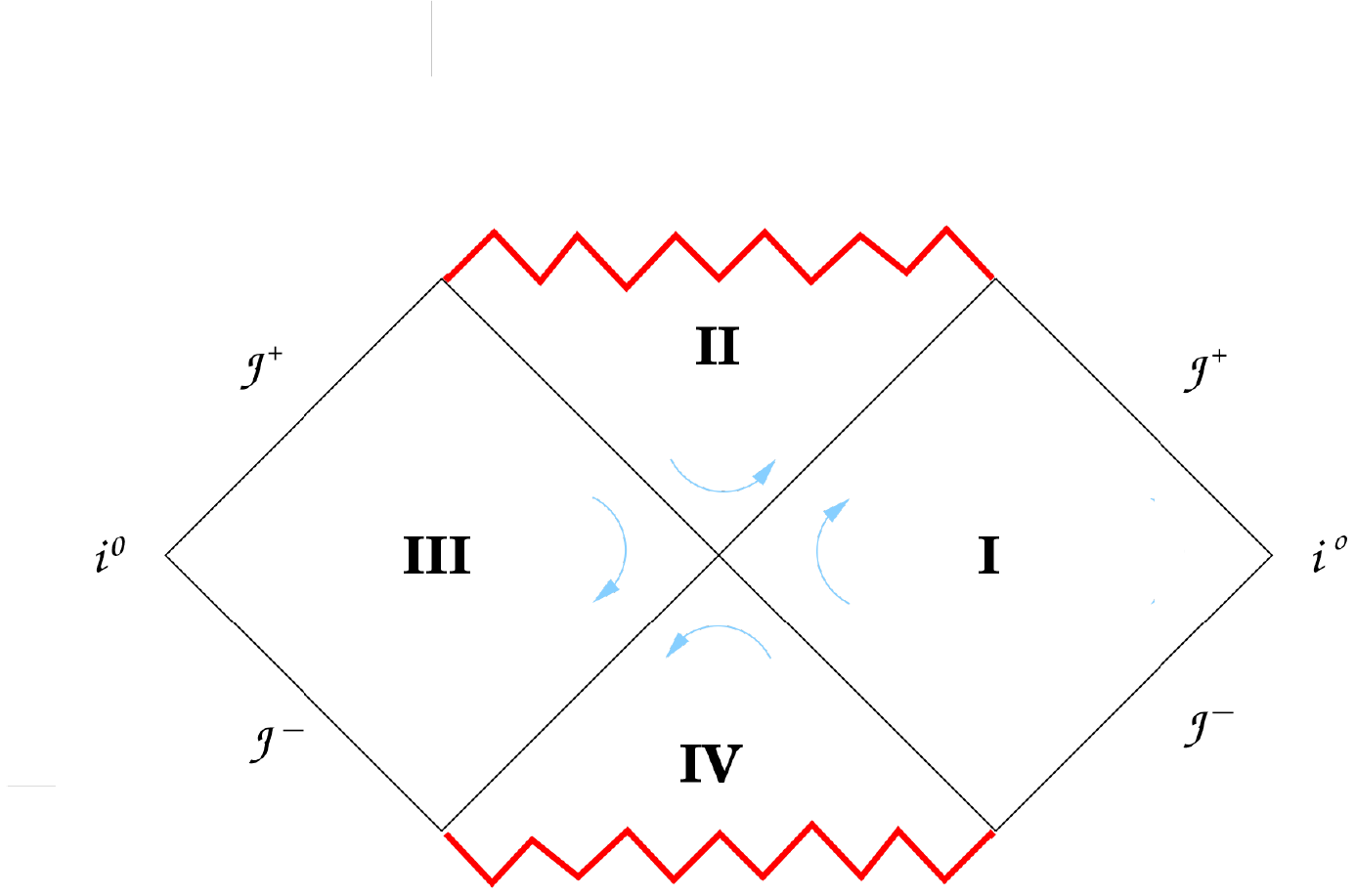} \\ (a)
         \end{center}
    \end{minipage}
   \hspace{.7in}
    \begin{minipage}{1.7in}
      \begin{center}
      \includegraphics[width=1.6in,height=1.6in]{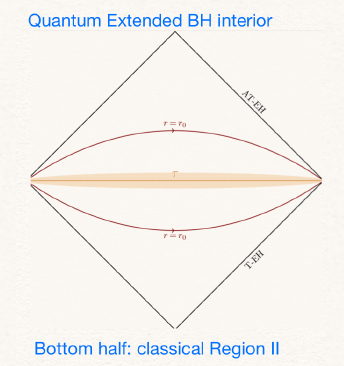} \\ {(b)}
       \end{center} 
   \end{minipage}
      \hspace{.5in}
       \begin{minipage}{1.7in}
        \begin{center}
         \includegraphics[width=1.6in,height=1.6in]{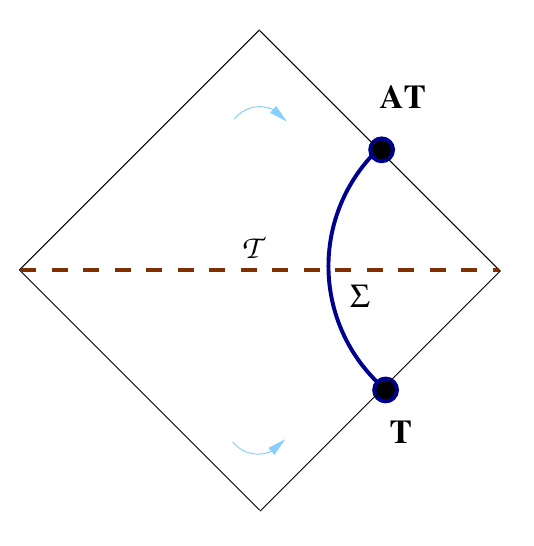} \\ {(c)}
          \end{center}
  \end{minipage}
\caption{\footnotesize{ (a) \emph{Left Panel}: Penrose diagram of Kruskal space-time. Only region II is of interest for the issue of singularity resolution.
(b) \emph{Middle Panel}: The LQG extension of region II of Kruskal space-time via singularity resolution. Singularity is replaced by a regular 3-manifold $\tau$ that separates the trapped and anti-trapped regions.
(c) \emph{Right Panel}: A time-like 3-surface $\Sigma$ joins two MTSs (depicted by blobs), one on the Trapping horizon that constitutes the past boundary of the trapped region and the other on the Anti-Trapping horizon that constitutes the future boundary of the anti-trapped region.}}
\vskip-0.5cm
\label{fig:3}
\end{center}
\end{figure}

This long detour into LQC may seem like a major digression in the present context of black holes. But in fact it is directly relevant to the issue of what happens to the black hole singularity. For, the region II of the Kruskal space-time that contains this singularity (see Panel (a) in Fig. \ref{fig:3}) is isometric to a homogeneous cosmology: the vacuum Kantowski-Sachs model. This region is foliated by the $r={\rm const}$ space-like surfaces each of which constitutes an orbit of the 4 Killing fields of the Kruskal metric. The coordinate $r$ plays the role of time and runs from the $r=2GM \equiv 2m$ at the past boundary that constitutes an isolated horizon (IH) to $r=0$ at the future boundary that represents a Big Crunch singularity of the Kantowski-Sachs model. Therefore, one can use methods used in LQC. There is rich literature on the subject and results based on effective equations have provided significant insights into what replaces the singularity and on the nature of quantum corrected geometry in the Planck regime, especially over the last $\sim\, 5$ years. For definiteness, we will focus on the first of these recent investigations \cite{aoslett,aos,ao} that set the stage and in which quantitative details have been worked out. (In other  recent investigations \cite{gop,mhhl} the effective equations are obtained using a more systematic approach but the final results are very similar.) Recall that, although the Kruskal space-time represents an idealized `eternal black hole', in the early investigations of the Hawking effect it provided valuable guidance on what to expect in more interesting collapsing situations. 
The situation is the same in LQG. Specifically, the following key results on the singularity resolution in Kruskal space-time have shaped our expectations in more realistic situations in which a macroscopic black hole  (i.e. one with $m \gg \lp$)  forms through gravitational collapse and then evaporates due to quantum radiation:
\vskip0.05cm

\emph{(i)} For macroscopic Kruskal black holes, the quantum geometry corrections are completely negligible near the non-expanding horizon (NEH) that constitutes the past boundary of region II in Panel (a) of Fig. \ref{fig:3}. For example, for a solar mass black hole, they are of the order of 1 part in $10^{115}$! However, they grow as $r$ decreases, leading to the resolution of the classical singularity at $r=0$. In the quantum corrected effective geometry, the singularity is replaced by a regular \emph{transition surface} $\tau$. This a space-like 3-manifold which is again an orbit of the 4 Killing fields and therefore a homogeneous 3-manifold foliated by round 2-spheres. The radius of these 2-spheres is given by $\rm{Rad}_{\tau} \approx (\uDelta^2 m)^{\f{1}{3}}\times 10^{-2}$. (By inspection, $\rm{Rad}_{\tau}$ goes to zero in the limit the area gap $\uDelta$ goes to zero, i.e., the limit in which quantum geometry effects vanish and $\tau$ coincides with the classical singularity, just as one would expect.) In the quantum corrected geometry, then,  $r$ ranges from $r \approx 2m$ on the past NEH to $r= \rm{Rad}_{\tau}$. \vskip0.05cm
\emph{(ii)} The quantum corrected geometry is smooth in this region and, as shown in the Panel (b) of Fig. \ref{fig:3}, effective equations extend it to a smooth metric the future of the transition surface $\tau$, all the way till a future IH.  As one would expect from Kruskal geometry, to the past of $\tau$ one has a trapped region: the expansion of both null normals to round 2-spheres is negative. Interestingly, \emph{they both vanish on $\tau$}! Thus the geometry at the transition surface is truly exceptional. To the future of $\tau$ both expansions are positive. Thus, $\tau$ is a boundary that separates a trapped region that lies to its past and an anti-trapped region that lies to its future. The radii of the round 2-spheres decrease monotonically as one moves from the past boundary of the Panel (b) of Fig. \ref{fig:3} to the transition surface $\tau$, and then increase monotonically as one moves to the future boundary. They acquire their minimum radius $\rm{Rad}_{\tau} \approx ({\uDelta}^2 m)^{\f{1}{3}}\times 10^{-2}$ on $\tau$ which grows as $m^{1/3}$ as the black hole mass grows. The past boundary is a trapping NEH, while future boundary, an anti-trapping NEH.
\vskip0.05cm
\emph{(iii)} Curvature of the quantum corrected metric is bounded throughout the diamond shown in Panel (b) of Fig. \ref{fig:3}. As one would expect, it reaches its maximum at the transition surface. Specifically, for the Kretschmann scalar:
\be \label{bounds}
 \mathcal{K}\mid_{\T}\,\,  \equiv \,\, R_{abcd}R^{abcd}\mid_{\T}\,\,=\, \frac{k_\circ}{\uDelta^2}\ + {\cal O}\big(({\uDelta}/{m^2})^{1/3} \,\ln\, ({m^2}/{\uDelta})\big)
\ee
where $k_\circ$ is a constant.  Thus, the classical singularity is naturally resolved, thanks to quantum geometry that provide us with a non-zero area gap $\uDelta$: In the classical limit, $\uDelta \to 0$,\, $\tau$ is replaced by the $r=0$ surface and we return to the Kruskal singularity. Note that the leading term in (\ref{bounds}) is \emph{universal}: It does not depend on the black hole mass. This is also what happens at the big bounce (that replaces the Big Bang) in quantum cosmology: the curvature at the bounce provides a universal upper bound in Friedmann-Lema\^itre models.  \vskip0.05cm
\emph{(iv)} As one moves away from $\T$, these curvature scalars rapidly approach their classical values even for very small microscopic black holes. For instance, while the horizon radius of the effective solution is always larger than that of its classical counterpart, even for $m=10^4 \lp$, the relative difference is $\sim 10^{-15}$ and, as remarked already, for a solar mass black hole, it is $\sim 10^{-115}$! Finally one can ask for the relation between the radius $r_{{}_{\rm T}}$ of the trapping horizon, and the radius  $r_{\rm AT}$ of the anti-trapping horizon. Are they approximately the same? The answer is in the affirmative for macroscopic black holes, even though the `bounce' is not exactly symmetric. For a stellar mass black hole for example, $r_{{}_{\rm T}} = 3$km and $r_{\rm AT} = 3\,(1 + {O}(10^{-25}))$km. \vskip0.05cm
\emph{(v)} What happens to the singularity theorems? Indeed, this question was raised in the context of a singularity resolution after Adrian del Rio's talk at the BHIO conference. The answer is that, as in LQC, they are bypassed because quantum corrections to  Einstein's equations become significant as one approaches $\tau$. More precisely, the Ricci tensor of the quantum corrected metric is non-zero and one can use it to define an `effective stress-energy tensor' $T^{\rm eff}_{ab}$ via $G_{ab} = 8\pi G T^{\rm eff}_{ab}$. As one would expect, $T^{\rm eff}_{ab}$ fails to satisfy even the weak energy condition. Except very near the two horizons, the energy density is negative. For a black hole with $m = 10^6 \lp$, it  becomes $\mathcal{O}(10^{-1})$ in Planck units at the transition surface $\tau$ but  decays very rapidly as moves away from $\tau$ and is $\mathcal{O}(10^{-20})$ near the two horizons. The square root of the Kretchmann scalar (that has the same dimensions as the energy density) is $\mathcal{O}(10^{-12})$; thus the contribution of the Ricci curvature to the total curvature is negligible even for these very small macroscopic black holes. 
\vskip0.05cm
\emph{(vi)} Several non-trivial checks have been made on this geometry to verify overall consistency. A conceptually interesting one comes from the Komar mass associated with the translation Killing field $\partial/\partial t$. Consider the values of the Komar mass evaluated on a 2-sphere on the trapping horizon and another 2-sphere on anti-trapping horizon, the two being connected by a 3-manifold $\Sigma$ as in Panel (c) of Fig. \ref{fig:3}. Now, in the classical theory, the Komar mass $M_{\rm K}$ defined by the translational Killing field is given by (half the) horizon radius. As we saw, for macroscopic black holes the radii $r_{{}_{\rm T}}$ and  $r_{{}_{\rm AT}}$ are essentially the same. On the other hand, the difference between the Komar mass evaluated at the anti-trapping horizon and the trapping horizon is given by the integral involving stress-energy tensor over a 3-manifold $\Sigma$ joining cross-sections of the two horizons (see Panel (c) of Fig.~\ref{fig:3}),
\be \label{balance2} M_{\rm K}^{\rm AT} - M_{\rm K}^{\,\rm T}\, =\, 2\,\int_{\Sigma}\! \big(T_{ab}^{\rm eff}\, -\, \f{1}{2} T^{\rm eff}\,\, g_{ab}^{\rm eff}\big)\, X^a \rmd \Sigma^b \, ,\ee
and for macroscopic black holes the integrand of the right is \emph{large and negative} near $\T$ (because it represents the effective energy density). How can the two Komar masses be the same, then?  It turns out that the integrand of (\ref{balance2}) is indeed large and negative for macroscopic black holes, but its numerical value is very close to $-2M_{K}^{\rm T}$. Therefore the Komar mass associated with the anti-trapping horizon is given by $M_{\rm K}^{\rm AT} \approx M_{\rm K}^{T} - 2M_{K}^{\rm T} = -M_{\rm K}^{\rm T}$, and the minus sign is just right because while the translational Killing field is {\rm future} directed at the 2-sphere on the trapping horizon {T}, it is \emph{past} directed on 2-sphere on the anti-trapping horizon {AT}! This resolution is an example of the conceptually subtle nature of the quantum geometry in the diamond bounded by the two horizons.\vskip0.1cm

We will conclude this subsection with two remarks:

1. Since the black hole singularity lies in region II of Kruskal space-time, in the above discussion we focused on the quantum extension of this region. The resulting quantum corrected (or effective) metric is well-defined on the boundaries: the trapping NEH and the anti-trapping NEH. It is then natural to ask if we can extend the metric beyond these boundaries in a systematic manner to asymptotic regions. This is indeed possible (see in particular \cite{mhhl,gop}).   

2. The quantum extensions of the type depicted in Panel (b) of Fig. \ref{fig:3} is often referred to as representing a ``black hole to white hole transition" because one has a trapped region to the past of $\tau$ and an anti-trapped region to the future. We have avoided this terminology because it has other connotations that are not realized. In particular, one loses predictivity in presence of white holes since anything can come out of their singularity. In the LQG transition from trapped to anti-trapped regions, on the other hand, there is no singularity and physics is completely deterministic across $\tau$ that replaces the singularity.   
\vskip0.1cm

Together with the discussion of DHs of section \ref{s2.1}, results summarized in this subsection have provided considerable intuition, streamlining possibilities both for permissible quantum geometries and for pathways to the recovery of information in the black hole evaporation process. These two sets of concrete results are used as stepping stones in current LQG investigations aimed at obtaining a complete description of the evaporation process. In the next section, we will summarize a mainstream perspective that has resulted.

\section{Black hole evaporation in LQG}
\label{s3}

Let us now turn to the issues related to the dynamics of black hole formation and evaporation in quantum theory using results of section \ref{s2} as guidelines. We will divide this discussion into two parts. In the first, we will focus on the semi-classical sector that excludes the Planck regime, and in the second we will discuss evolution to its future through the Planck regime. We will discuss the two interrelated but rather distinct issues: (i) Nature of the quantum corrected space-time geometry; and, (ii) issues related to entanglement, von-Neumann entropy and purity of the final state at $\scrip$. Of course, important issues remain, especially in the second part. Nonetheless, the hope is that this streamlining of possibilities will lead to focused efforts to weed out ideas and concentrate on the viable paths that remain.

\subsection{The semi-classical regime}
\label{s3.1}
 
To ensure the validity of the semi-classical approximation, in most of this sub-section we will consider a solar mass $M_\odot$ blackhole that is formed by a gravitational collapse and let it evaporate till it reaches the lunar mass $M_{\text{\leftmoon}}$. This process takes some $10^{64}$ years. But even at the end of this long evaporation time, the final black hole has a macroscopic mass. Therefore, during this entire phase of evaporation, the process should be well approximated by semi-classical gravity. While there have been proposals advocating large deviations from the semi-classical theory even for astrophysical black holes (e.g. due to `firewalls'), in light of the results of the LIGO-Virgo collaboration, these proposals are no longer regarded as viable by most of the community. (For a general discussion on implausibility of the failure of semi-classical gravity well away from the Planck regime, see in particular \cite{ori1}.)

As explained in footnote \ref{fn1}, to discuss the issue of information-loss/unitarity in a meaningful way, one needs to work with a closed system. Therefore, we will focus on the system consisting of massless scalar fields $f$ coupled to gravity in 4-dimensions. At the classical level, we will use Einstein's equations (which, however, will be appropriately modified in the quantum theory). We will restrict ourselves to spherical symmetry and, for simplicity ignore back scattering as in Varadarajan's model \cite{mv} (see the last para in section \ref{s2.1}). Finally, to justify the semi-classical approximation --in which one treats matter quantum mechanically but ignores the quantum fluctuations of geometry-- we will use a large number $N$ of scalar fields and work with the $1/N$ expansion as in section \ref{s2.2}. 

\begin{figure} 
 
  \begin{center}
    \begin{minipage}{1.5in}
      \begin{center}
        \includegraphics[width=1.5in,height=2.7in,angle=0]{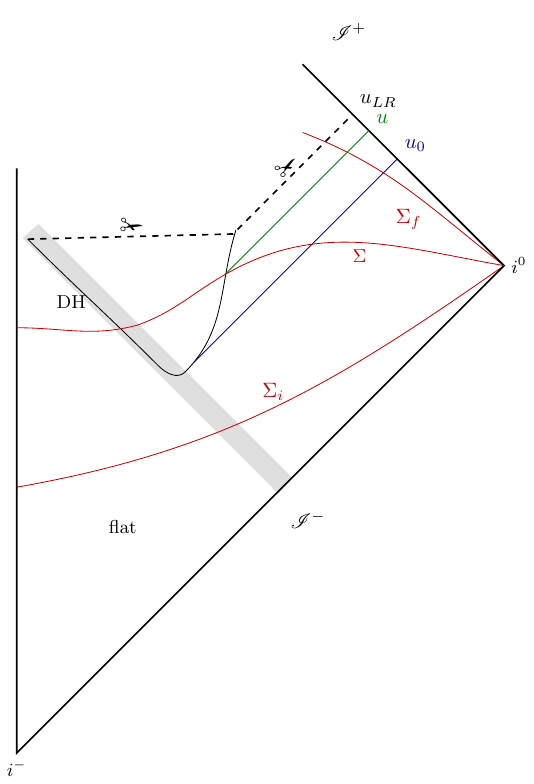} \\ (a)
         \end{center}
    \end{minipage}
   \hspace{.7in}
    \begin{minipage}{1.7in}
      \begin{center}
      \includegraphics[width=1.3in,height=2.7in,angle=0]{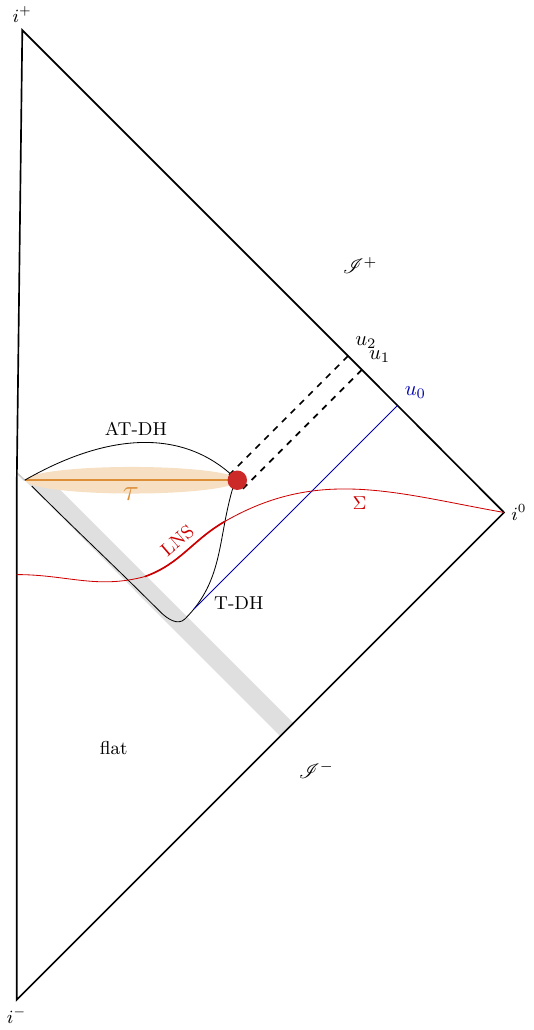} \\ {(b)}
       \end{center} 
   \end{minipage}
      \hspace{.5in}
       \begin{minipage}{1.7in}
        \begin{center}
         \includegraphics[width=1.4in,height=2.7in,angle=0]{TDL-traditional-eva.pdf} \\ {(c)}
          \end{center}
  \end{minipage}
\caption{\footnotesize{ (a) \emph{Left Panel}: The expected semi-classical space-time in LQG. Planck regime has been cut out. The DH has two branches, the expanding, space-like, left branch is formed as a result of the collapse and the contracting, time-like, right branch (that replaces the event horizon of the Panel (c) of Fig. \ref{fig:1}). Each branch is a \TDH\!\!. Together, they bound a trapped region.\,\, (b) \emph{Middle Panel}: The conjectured full space-time of LQG. Curvature is Planck scale in the shaded pink region that contains the transition surface $\tau$. To the past of $\tau$ we have a 
trapped region, bounded by a \TDH and the transition surface $\tau$, and to the future of which we have an anti-trapped region, bounded  by $\tau$ and an  \ATDH\!\!. The well-developed approximation methods of LQG are inapplicable to the prink blob where the fluctuations of geometry could be very large. Because their influence has not been explored, the space-time portion to the future of the \ATDH is left blank.\,\,
(c) \emph{Right Panel}: Hawking's original Penrose diagram for an evaporating black hole is reproduced here for ready comparison with the LQG proposals.}}
\vskip-0.5cm
\label{fig:4}
\end{center}
\end{figure}

In the incoming state on $\scrim$, one of the quantum scalar fields, say $\hat{f}_1$, will be assumed to be in a coherent state that is peaked at a classical scalar field $f_1^\circ$  (of compact support on $\scrim$) that undergoes a prompt collapse to form a black hole. The remaining $N-1$ quantum fields $\h{f}_i$ will assumed to be in their vacuum state (as in \cite{apr,mv}). The semi-classical Einstein's equations governing this system are:
\be \label{semi-class} G_{ab}^{\rm (sc)} = 8\pi G_{\rm N}\,  \langle\, \hat{T}_{ab}\, \rangle_{\rm ren} \quad {\rm and} \quad  \Box\, \hat{f_i} = 0\, , \hbox{\rm{with} $I = 1,\ldots N$} \ee
where $G_{ab}^{\rm (sc)}$ is the Einstein tensor of the semi-classical metric $g_{ab}^{\rm (sc)}$ and the expectation value of the renormalized stress-energy tensor is computed using the Heisenberg state $\Psi$ and the space-time metric $g_{ab}^{\rm (sc)}$. The metric $g_{ab}^{\rm (sc)}$ does include quantum corrections but they are induced only by quantum matter (since the quantum geometry terms  induced by the area gap are completely negligible away from the Planck regime). These corrections to geometry are adiabatic and small. But the infalling negative energy flux has a non-trivial effect on the horizon structure already at the start of the evaporation process. In the classical theory, the space-like dynamical horizon DH would have continuously joined on to an isolated horizon --the future part of the event horizon (see the Panel (c) of Fig. \ref{fig:1}). Now, this space-like branch of the DH turns around and becomes time-like (see Panel (a) of Fig. \ref{fig:4}). Both branches are trapping DHs (T-DHs). Together, the two branches of the DH enclose a trapped region: in this region expansions of both bull normals to the MTSs are negative. 

During the evaporation process, modes are created in pairs. One escapes to $\scrip$ and its partner is trapped in this region. Therefore, as the black hole evaporates the total state on a Cauchy surface (such as $\Sigma$ in Panel (a) of Fig.\ref{fig:4}) is increasingly entangled. Now, because the right branch of the DH is time-like, light \emph{can} escape the trapped region (in sharp contrast to the situation where the trapped region is bounded by an event horizon). Therefore, one might imagine that information could leak out from the trapped region, leading to purification of the state at $\scrip$ already in the semi-classical regime \cite{hayward-conf}. But, as we already indicated in section \ref{s2.1.2}, a careful analysis of the partners modes that go out to $\scrip$ shows that correlation will not be restored at $\scrip$ during the semi-classical phase \cite{agulloetal}. Thus, even at the end of the evaporation process now under consideration, when the black hole has shrunk from solar $M_\odot$ to lunar mass $M_{\text{\leftmoon}}$, the quantum state remains entangled.

But then there is an apparent paradox already in the semi-classical regime. Since $M_{\text{\leftmoon}} \sim  10^{-7}\, M_\odot$, at the end of this long evaporation process most of the initial mass is carried away to $\scrip$ by the outdoing modes. A back of the envelope calculation shows that a very large number $\mathcal{N}$\, ($\sim 10^{75}$)\, of quanta escape to $\scrip$ and all of them are correlated with the ones that are trapped in the region enclosed by \TDH\!\!\!. Therefore, at the end of the semi-classical process under consideration, one would have a huge number $\mathcal{N}$ of quanta both at $\scrip$ and in the trapped region. But the mass associated with the trapped region is only $10^{-7}$ times that carried away to $\scrip$. Furthermore, the radius of the outer part of \TDH has shrunk to only $0.1$mm -- the Schwarzschild radius of a lunar mass black hole. How can a \TDH with just a $0.1$mm radius accommodate all these $\mathcal{N}$ modes? Even if we allowed each mode to have the (apparently maximum) wavelength of $0.1$mm, one would need the horizon to have a huge mass --some $10^{22}$ times the lunar mass!  While these considerations are quite heuristic, one needs to face the conceptual tension: At the end of the semi-classical phase under consideration, the trapped region seems to have ``too many quanta to accommodate, with a tiny energy budget". 

The resolution of this apparent paradox lies in the fact that the geometry of the trapped region has some rather extraordinary features that had not been noticed until relatively recently. Recall that the evaporation process is extremely slow, and the \TDH mass of the time-like branch decreases in response to the outgoing flux by the the standard Hawking formula: $\rmd M_{\TDH}/\rmd v = -  \hslash/(GM_{\TDH})^{2}$, where $v$ is the retarded time coordinate as in Hawking's calculation. Using these two features, it has been argued that the metric in the trapped region should be well-approximated by that of the Vaidya space-time in which the mass $M(v)$ decreases from $M_\odot$ to $M_{\text{\leftmoon}}$. To probe this geometry, let us foliate the trapped region by 3-dimensional space-like surfaces, each with topology $\mathbb{S}^2\times \mathbb{R}$. There are two particularly natural choices: $\mathcal{K} = {\rm const}$ surfaces, and $r = {\rm const}$ surfaces, where $\mathcal{K}$ denotes the Kretchmann scalar and $r$ denotes the radius of the round 2-spheres $\mathbb{S}^2$. The radius of the time-like portion of $\TDH$ decreases from $3$km to $0.1$mm. But, as a simple calculation shows, the 3-dimensional surfaces develop \emph{astronomically long `necks'} along the $\mathbb{R}$-direction. By the time we have reached lunar mass $M_{\text{\leftmoon}}$, the lengths of these necks $\ell_{N}$ are given by:  $\ell_{N} \approx 10^{64}$ \emph{light years} for the first foliation, and $\ell_{N} \approx 10^{62}$ \emph{light years} for the second \cite{aaao,aa-ilqg}!%
\footnote{one can also consider $\rm{tr} K = {\rm const}$ slices \cite{cdl}. But they do not provide a foliation of the entire trapped region  But the main phenomenon of developing enormously long necks,  occurs also on these 3-surfaces.} 
These astronomically large lengths can result because the process has a really huge time at its disposal; $10^{64}$ years is $\sim 10^{53}$ times the time elapsed since nucleosynthesis. 

This enormous stretching is analogous to the expansion that the universe undergoes in (an anisotropic) cosmology. Recall that during the cosmic expansion --e.g. during inflation-- the wavelengths of modes get stretched enormously. This suggests that partner modes in trapped region will also get enormously stretched during evolution and become infrared. Can this phenomenon resolve the quandary of `so many quanta with so little energy'? At the qualitative level of this discussion, the answer is yes. With such infrared wavelengths, it is easy to accommodate $\mathcal{N}$ modes in the trapped region with the energy budget only of $M_{\text{\leftmoon}}$. Thus, even though the  outgoing modes carry away almost all of the initial mass  $M_{\odot}$ to $\scrip$, there is no obstruction to housing all their partners in the trapped portion of the slice $\Sigma$ of Panel (a) of Fig.~\ref{fig:4} with the small energy budget of just $10^{-7}M_{\odot}$.  This argument removes the necessity of starting purification by Page time. In the LQG perspective, purification occurs at a much later stage.

To summarize, in the long semi-classical phase the outgoing modes that register on $\scrip$ are entangled with their partners, confined to the trapped region. But the outgoing quanta carry away most of the total energy $M_\odot$ leaving only a small remainder $M_{\text{\leftmoon}} \approx 10^{-7} M_\odot$ in the trapped region. This occurs because there is a steady negative energy flux going into the trapped region that erases most of the $M_\odot$ of infalling energy, just as one would expect from energy conservation. But energy considerations are distinct from the entanglement issues and this raises an apparent paradox already in the semi-classical phase: How can the small energy budget of the trapped partner modes suffice to hold `as many' of them as those went out to $\scrip$ carrying huge total energy $(1-10^{-7})\, M_\odot$? The resolution lies in the fact that the space-time geometry of the trapped region is highly non-trivial: As evaporation proceeds these modes get stretched \emph{enormously} and become infrared. Therefore entanglement between the modes that have registered at $\scrip$ and those in the trapped region can persist, even though the energy associated with the two sets is vastly different. Thus, at the end of the semi-classical phase, if one were to trace over the trapped partner modes, the state at $\scrip$ would be mixed. This but to be expected because the semi-classical $\scrip$ is not complete. Whether purity is finally resorted at $\scrip$ depends on what happens to the future of the semi-classical region. 

\subsection{Evolution through the Plank regime}
\label{s3.2}

In the last subsection we considered the phase during which a solar mass black hole shrinks to lunar mass to make it obvious that we are in the semi-classical regime. But, as is widely expected, the semi-classical approximation should continue to be viable until the curvature at the \TDH is $\sim\, 10^{-6}$ in Planck units, i.e., the black hole has shrunk to $\sim 10^3$ Planck mass. However, then one enters the Planck regime --shown in pink in Panel (b) of Fig. \ref{fig:4}-- where one has to use the full quantum state $\Psi$ of matter \emph{and} geometry and study its evolution. The key question that remains is: What happens to quantum geometry \emph{and} the infrared partner modes of quantum matter that are confined to the trapped region of the semi-classical phase? Since they are entangled with the modes that went out to $\scrip$ throughout the very long evaporation process in which the $M_\odot$ black holes shrinks to $\sim\,10^3$ Planck mass, the associated von-Neumann entanglement entropy is very large. Nonetheless,  there \emph{is a pathway to restoration of purity} of the full quantum state at $\scrip$: If the partner modes were to evolve across the Planck regime that lies to the future of the semi-classical phase, they would arrive at $\scrip$ where they would be correlated with the modes that arrived at early times, just as they were while in the trapped region. Thus, at the end of the process, the complete quantum state at $\scrip$ should be pure, just as the complete quantum state was pure on a Cauchy surface $\Sigma$ passing through the semi-classical region (see Panel (b) of Fig.\ref{fig:4}).

To find out what happens to the partner modes in the Planck regime, we have the very difficult task of 
evolving quantum fields $\h{f}_i$ on the quantum geometry in this region. There are two aspects to the difficulty: (i) one cannot ignore the quantum fluctuations of geometry because we have Planck scale curvature, and, (ii) dynamical time scales for changes in the matter field and geometry can be Planckian, so that the process is highly non-adiabatic. Now in the long shaded, pink region of Panel (b) depicting the Planck regime, we do face the first issue. However, for a \emph{very} long interval in the advanced coordinate $v$, the evaporation process is so slow that the adiabatic approximation holds. Therefore, we can divide this region from left end to right into large intervals in each of which the geometry does not change significantly. This approximation fails at the right end of the region, depicted by the red blob. Here, the time-like branch of \TDH  has mass less than $\sim 10^{3}$ in Planck units, whence one expects dynamical time scales to be also Planck scale. Here one faces the two difficulties simultaneously.

Let us therefore postpone the discussion pertaining to this red blob and first consider the rest of the Planck regime to its left, where we have Planck scale curvature but the adiabatic approximation holds. From our discussion of section \ref{s3.1}, one expects this region to be very long but foliated by very small 2-spheres. Fortunately, prior experience in LQC --in particular the detailed investigation of the  propagation of cosmological perturbations on the \emph{quantum} FLRW geometries-- suggests a strategy to analyze dynamics in this region (see, e.g., section II.C of \cite{aan3}). Specifically, although the quantum state of geometry does have large fluctuations, dynamics of the scalar fields $\h{f}$ is not sensitive to all of them. Consequently, one can construct from the quantum state of geometry a smooth metric $\t{g}_{ab}$ that knows not only the expectation value of the metric operator but also those fluctuations in geometry that dynamics of quantum fields $\hat{f}_i$ is sensitive to. %(Thus, it has more information about the full quantum state of geometry than the effective metric, discussed in section \ref{s2.2}, that knows only the expectation value.)
$\t{g}_{ab}$ is called the \emph{dressed metric} and by construction its coefficients depend on $\hbar$. The difficult task of evolving quantum fields $\hat{f}_i$ on quantum geometry is reduced to that of evolving them on the space-time of the dressed metric $\t{g}_{ab}$. We will assume that the  $\t{g}_{ab}$ can be found in the adiabatic phase of the Planck regime in the present case as well.

Results reported in Section \ref{s2.2} on geometry in the Planck regime suggest that the shaded (pink) region will contain a transition surface $\tau$  (w.r.t. $\t{g}_{ab}$) that replaces the classical singularity and separates the trapped region that lies to its past from the untapped region that lies to its future. The metric $\t{g}_{ab}$ is expected to capture three distinct effects: (i) those that originate from quantum geometry, originating in the area gap $\uDelta$ (as in Section \ref{s2.2}); (ii) those that are induced on $\t{g}_{ab}$ by the falling quantum matter in the incident pulse of the scalar field at the left end of the (pink) shaded region, and, (iii) those associated with the negative energy flux into the trapped region across the time-like part of the \TDH\!\!\!. Results reported in \ref{s2.2} strongly suggest that the first set of effects will decay rapidly away from the pink region with Planck curvature, so that in the semi-classical region  $\t{g}_{ab}$ will be well approximated by $g^{\rm (sc)}_{ab}$ used there in section \ref{s3.1} ensuring consistency with semi-classical considerations.  
%Results reported in \ref{s2.2} strongly suggest that the first set of effects will decay rapidly as we move away from Planck curvature into the semi-classical region . Therefore in the semi-classical region, $\t{g}_{ab}$ will be well approximated by $g^{\rm (sc)}_{ab}$ used there. 
As we move to the future of the (pink) shaded region, one would encounter an anti-trapping dynamical horizon \ATDH (see Panel (b) of Fig.~\ref{fig:4}). The region enclosed by the transition surface $\tau$ to the past and \ATDH to the future would be anti-trapped with respect to $\t{g}_{ab}$. While geometry in the region bounded by \TDH to the past and \ATDH to the future is qualitatively similar to that of the quantum extension of Kruskal space-time (Panel (b) of Fig.~\ref{fig:3}), there is a key difference because we are now in a dynamical situation. While the boundaries in the Kruskal extension are null IHs, now the past boundary is a \TDH with a space-like and a time-like branch, and the future boundary is an \ATDH that is space-like.

Finally, one would expect that the region to the future of \ATDH would be well approximated by an approximately flat metric with an outgoing flux with a small total energy ($\sim 10^{3}$ in Planck units $\sim 10^{-2} gm$) spread over astronomical scales. It will describe the propagation of the infrared modes that will emerge from the \ATDH and arrive at $\scrip$ at very late times.  Recall that these are the partner modes which were entangled with the outgoing modes that carried away most of the initial ADM mass to $\scrip$. In the LQG scenario, then, correlations are finally restored at $\scrip$ where, in the end, the partner modes also arrive. The total energy carried by the two sets of modes is very different. But this is not an obstruction for restoring correlations, i.e., for the `purification' to occur since, as emphasized before, there is no direct correlation between energy flux and entanglement.

A commonly held notion that purification should occur before Page time (when the black hole has lost only half its mass through quantum radiation) implies that correlations have to be restored already in the semi-classical phase. As discussed earlier, recent investigations provide strong arguments against this possibility \cite{agulloetal}. In the LQG scenario purification occurs \emph{much later} because singularity resolution allows the partner modes to emerge from the trapped region and reach $\scrip$, traveling across the transition surface $\tau$ that replaces the singularity. The timescale of this purification process would be very long, $\mathcal{O}(M^{4})$ \cite{ori4,ebtdlms,bcdhr}. But there is a pathway to restoration of correlations at $\scrip$, thereby making the total quantum state at $\scrip$ pure. This is a mainstream viewpoint in LQG. 

However, a rather glaring open issue remains: the red blob in Panel (b) of Fig. \ref{fig:4}. In this region,  not only is the curvature of Planck scale, but it is varying extremely rapidly because it lies at the end point of the evaporation process. Together, these two effects make the known approximation methods inapplicable. There \emph{are} approaches to evolve across this region using full quantum gravity both in the Hamiltonian \cite{ori5} and the path integral (spinfoam) approaches  \cite{dhrv}. But they have limitations. So far they do not account for the very non-trivial features that arise in the pathway to restoration of purity outlined above. Let me use Panel (b) of Fig. \ref{fig:4} to illustrate the type of challenges that remain. If this scenario of information recovery is correct, then one would expect that, as one approaches $u=u_1$ along $\scrip$, the temperature of radiation would grow since the time-like potion of \TDH is rapidly shrinking to Planck size. Therefore, the emitted quanta would be ultra-violet at $\scrip$. On the other hand, to the future of $u=u_2$ on $\scrip$, one would find infra-red modes. How does this dramatic transition from ultraviolet to infrared come about? Presumably this is because of  non-trivial effects originating in the red blob. But so far we do not have a systematic understanding of how this would come about. Similarly, there are issues concerning the anti-trapping horizon \ATDH\!\!\!.  Is the \ATDH stable w.r.t. small perturbations? 
\footnote{The trapping horizon \TDH should be stable because both its branches are so-called FOTHs --future outer trapping horizons \cite{hayward,ak2}. The \ATDH on the other hand is not; it is `inner' in this terminology.}
One could presumably address this issue using the metric $\t{g}_{ab}$ but it is not clear what the class of physically relevant perturbations would be. A second issue concerns the geometry in its future neighborhood of \ATDH away from the red blob. In our scenario, the \ATDH  is foliated by 2-spheres with radii less than $\sim\, 10^3 \lp$ but the length in the transverse direction is enormous, $> \, 10^{60}$lyrs (both measured using $\t{g}_{ab}$). Since the total energy density in the infra-red modes is small and spread out over astronomical length scales, the geometry to the future of \ATDH should be approximately flat (assuming the red blob has no effect on it). Therefore if one chose an approximately flat space-like 3-surface $\Sigma$ in a neighborhood of \ATDH to its future and adapted to spherical symmetry, it would be foliated by 2-spheres whose radii increase monotonically as we move to right. Therefore, say, half way between the left end and the red blob, the radius of these 2-spheres on $\Sigma$ would have to be $\sim 10^{60}$lyrs, while the radius of the `corresponding' 2-sphere on \ATDH would be $\sim 10^3 \lp$. Is there an admissible nearly flat metric that admits such dramatic growths in the size of 2-spheres as we pass from \ATDH to a nearby $\Sigma$?  I should add that I do not know of any concrete obstructions to constructing the required nearly flat 4-geometry, or showing stability of the \ATDH\!\!. In fact I think that general ideas underlying the pathway are likely to be correct. But such issues have not been adequately addressed yet and so there is considerable food for thought.

To summarize, LQG does provide a pathway to obtain a coherent space-time description of the black hole evaporation process in which correlations are restored at late time on $\scrip$, restoring the purity of the final state. The value of the pathway lies in the fact that one has a concrete scenario that one can try to prove or falsify. Along the way one may find genuine surprises. A priori there is a  possibility that genuine quantum gravity effects associated with the red blob may lead to information loss, e.g., by creation of a baby universe. Even though prior experience with the Planck scale regime --especially in LQC-- suggests that this is unlikely, this issue needs a much more careful scrutiny than it has received in the LQG community so far.
%ruled out the possibility that genuine quantum gravity effects associated with the red blob will lead to information loss, e.g., by creation of a baby universe  --not widely shared -- 
%the possibility that genuine quantum gravity effects associated with the red blob will lead to information loss, e.g., by creation of a baby universe is not ruled out. But prior experience with the Planck scale regime --especially in LQC-- leads me to believe that this is unlikely.

\section{Discussion}
\label{s4}

The last three sections summarized a mainstream LQG viewpoint on the process of black hole evaporation. As emphasized in sections \ref{s1} and \ref{s2}, it departs from commonly held views based on Hawking's original proposal in that there is neither an EH nor a singularity in the final picture \cite{aamb}. These features constitute corner stones of most of the LQG work on black hole evaporation. The detailed scenario for the evaporation process was then built using four sets of concrete results: (i) properties of DHs in classical and semi-classical gravity; (ii) the natural resolution of space-like singularities due to quantum geometry effects; (iii) properties of the quantum extension of the part II of Kruskal space-time that contains the black hole singularity; and, (iv) a strategy to handle quantum fields propagating on quantum space-times in the Planck regime when dynamics is adiabatic. These results were obtained by a very large number of researchers, and even with a rather long bibliography I could include only a sample of this rich literature. 

In the LQG community, there is a general agreement on the description semi-classical phase summarized in section \ref{s3.1}, although we need more detailed calculations to arrive at the space-time geometry in the trapped region directly from the semi-classical equations. Currently, much of our understanding is shaped by the detailed analyses of the CGHS model \cite{ori2,apr,ori5}. While it captures several features of the 4-d spherically symmetric gravitational collapse and subsequent evaporation, as discussed at the end of section \ref{s2.1.2}, it also differs from the 4-d model in certain important respects. However, a recently proposed model \cite{mv} does not have these limitations, even though it is also exactly soluble classically. Its semi-classical equations (in the $1/N$ expansion) have been written down and they are similar to those of the CGHS model. A high precision numerical study will soon be undertaken by Fethi Ramazanoglu and Semith Tuna. They should provide a much more reliable description of the semi-classical phase of the 4-d evaporation process. This model does make an approximation: it ignores the back scattering effects. If one makes the same approximation in the derivation of the Hawking effect to begin with, one misses the gray body factors which can be added subsequently. The hope is that the situation would not be significantly different also at the semi-classical level when the back reaction is included. 

A concrete strategy to go beyond the semi-classical approximation is sketched in section \ref{s3.2}.
It provides a plausible pathway to restoring correlations on $\scrip$, thereby ensuring the purity of the final quantum state there. A visual comparison between the LQG proposal depicted in Panel (b) of Fig. \ref{fig:4} and the commonly used proposal shown in Panel (c) brings out the fact that this pathway is possible precisely because: (i) the EH of (c) is replaced by DHs in (b), and, (ii) the singularity in (c) is replaced by a regular transition surface $\tau$. However, difficult issues remain in the quantum evolution especially beyond the semi-classical regime, and variations on the strategy presented in section \ref{s3.2} are also being pursued in the LQG community (see, e.g. \cite{perez}). In my view, the key open issue is the following: The well-developed approximation methods of LQG are inapplicable to the red blob in the Panel (b) (where the trapping and anti-trapping regions meet) because not only does it have Planck scale curvature but it is also highly dynamical. So far this issue has not received as much attention as it deserves. The space-time portion to the future of the \ATDH is purposely left blank in panel (b) because of the uncertainties on the influence of the red blob on the geometry in this region. Nonetheless, there is value in having a concrete paradigm --such as the one sketched in section \ref{s3.2}: it raises specific interesting questions that one may not have envisaged, thereby providing directions for further work that can confirm or falsify expectations. In particular, a classification of all spherically symmetric spacetimes that could result from singularity regularization is available in the literature \cite{rubio}. It can be used in LQG because it is based purely on geometrical considerations, without any assumptions on the underlying dynamics. If one could argue that one of these possibilities is realized in the `red blob' and the `blank region' of panel (b), a space-time description would be available for the entire evaporation process. If not, the structure of the `red blob' and its influence would have to be analyzed using full LQG, e.g., using a quantum transition along the lines of \cite{ori4}. In that case, a complete description of the evaporation process would involve regions in which physics cannot be described using a smooth continuum equipped with a pseudo-Riemannian metric.

We will conclude with a few general remarks:\vskip0.05cm

1. In section \ref{s2.2}, we saw that region II of Kruskal space-time admits an LQG extension in which the singularity is replaced by a transition surface (see Panels (a) and (b) of Fig. \ref{fig:3}). As we remarked at the end of that section, this quantum corrected geometry has been also been extended to include asymptotic regions \cite{aos,gop,mhhl}. One can therefore ask if the quantum corrected metric gives rise to effects that would be relevant astrophysically. Given the quantum corrections to the metric near the horizon are \emph{extremely} small for astrophysical black holes, one would expect that the answer to be in the negative. This has been borne out in detailed analyses of quasi-normal modes (see. e.g.,\cite{kunstatter,dco}).

2. The LQG literature on collapsing models is very rich (see, e.g., recent discussions in \cite{hhcr,bcdhr,pmdcr,pmd,lmyz,hrs,fhwe,cfwe}) and often draws on earlier works on regular black holes \cite{hayward,frolov,bardeen}. These  models have provided us with concrete possibilities for the quantum corrected geometries. However, generally they do not discuss issues related to entanglement between the modes radiated to $\scrip$ and their partner modes that are initially trapped, nor to pathways to purification. At times, regular black hole models have suggested incorrect avenues for information recovery \cite{hayward-conf}. Finally, most of these models focus on stellar collapse which does not constitute a closed system that is necessary to the discussion of purification. That is why these models were not discussed in detail in this brief report, even though they have provided many interesting insights. 

3. The discussion of section \ref{s3.1} shows that, in the LQG perspective, there is a major difference between a \emph{young} lunar mass black hole that just formed due to gravitational collapse, and an isolated, \emph{old} lunar mass black hole that has resulted due to quantum radiation, starting from a solar mass black hole that was formed some $10^{64}$ years ago. While their dynamical horizons will have the same radius, $\sim\,0.1$mm, and mass $M_{\TDH} = M_{\text{\leftmoon}}$, their  external environment as well as internal structure will be \emph{very} different. In the case of an old black hole, a very large number of quanta would have been emitted to $\scrip$ and their partner modes would be trapped in the region enclosed by \TDH\!\!\!.  Therefore, the area of the time-like branch of the \TDH would not be a good measure of the von-Neumann entanglement entropy for the old black hole. %In the LQG perspective, for both black holes, area is a measure of the number of the quantum geometry states of the horizon itself that can interact with those in the trapped region as well as those that are outside. 

4. It is often argued that there is a potential problem with the notion old black holes (and hence with the discussion of the semi-classical sector in section \ref{s3.1}): Because old black holes can have small energy but an enormous number of modes, it should be easy to produce them copiously in particle accelerators. But these arguments use only the conservation laws normally used in computing scattering amplitudes in particle physics. Old black holes, on the other hand,  have astronomically long necks with a very large number of infrared modes. They are hardly particle like remnants! It is hard to imagine how such configurations can be created on time scales of accelerator physics \cite{ori1}.

5. Input from the journal suggested that adding a comparison between LQG and string theory/holography would be useful to the audience. %Over the years several quite different proposals have come from the string theory community and I can only include a few highlights. First,
I can only include a few highlights since a number of quite different proposals have come from the string community. In particular,
the AdS/CFT conjecture has been used to argue that Schwarzschild-like singularities will not be resolved in string theory \cite{eh}. Then, the space-time diagram describing an evaporating black hole would be like in Hawking's original proposal (Figure 4 (c)) and one is led to invoke novel mechanisms for information recovery on the portion of $\scrip$ to the past of the null ray labeled $u_{EH}$. Over the years, there have been proposals of `quantum xerox machines', `firewalls', and potential mechanisms based on `fast scramblers'. Each attracted a great deal of attention when %they were 
{it was first made,} but these ideas appear to have faded by now. The presumed necessity of novel mechanisms also led to proposals that there would be large violations of semi-classical gravity in tame regions in which space-time curvature is far from the Planck scale. As I mentioned in section \ref{s3.1}, these ideas are no longer regarded as viable in the mainstream. By contrast, as summarized in sections \ref{s2} and \ref{s3}, the overall LQG perspective has remained the same over the last two decades; much of the effort has been devoted to systematically develop a paradigm \cite{aamb} that emphasizes singularity resolution, and replacement of event horizons by quasi-local ones. 

Until recently, much of the literature in string theory/holography focused on black holes in presence of a negative cosmological constant as a simplified mathematical context to probe what would happen in the asymptotically flat case of direct physical interest. However, generally these arguments make a strong use of the asymptotically anti-de Sitter boundary conditions. It is not clear that one can remove these restrictions without altering conclusions.  Literature in LQG has focused on asymptotically flat space-times. %However, 
{ More} recently, with the introduction of worm-holes and replica islands, there has been a conceptual shift { in the string community, and this work does use asymptotically flat boundary conditions.}
% and this part of the string community uses asymptotically flat boundary conditions. 
However, string theory plays no essential role in this analysis; one evaluates path integrals as one would in quantum general relativity. As I understand from discussions and correspondence with practitioners, the analysis is semi-classical and does not address the issue of the fate of classical singularities. The emphasis is on the Page curve, but the turn-around of the %Page 
curve that is found { at the Page time} does not refer to the entanglement entropy (which would continue to { grow} in the semi-classical regime) but to another notion entropy that is deemed better suited for certain purposes (e.g. experiments that would try to measure entropy of the Hawking radiation). The turn around is attributed to topology change. But it requires large non-locality linking the black hole and distant radiation, and there is controversy on the viability of { this} proposal \cite{sm3}. By contrast, singularity resolution has been a focal point in LQG. { Also, one} focuses on the entanglement entropy which continues to grow throughout the semi-classical region and purification occurs at very late times, when the partner modes that fell across (the time-like portion of) the dynamical horizon T-DH arrive at $\scrip$ after crossing the anti-trapping dynamical horizon AT-DH.

In terms of steadiness of the overall perspective, LQG is similar to the fuzzball approach in the string community \cite{sm1,sm2}. There are also conceptual similarities in that there are no singularities, nor event horizons. Classical singularity theorems are again evaded by quantum modifications of Einstein's equations. However, the origin of these modifications is very different: In the fuzzball picture they arise from stringy corrections to Einstein equations, branes, and winding numbers, while in LQG their origin can be traced back to the \emph{quantum} nature of {the} space-time geometry itself. Therefore, the space-time descriptions of the Hawking process is also quite different.

\section*{Acknowledgments}

Over the years, I have profited greatly from stimulating discussions and correspondence on black hole evaporation with a large number of colleagues. For the material included in this article, I would especially like to thank I. Agullo, A. Almheiri, E. Bianchi, T. De Lorenzo, M. Han, B. Krishnan, D. Marolf, S. Mathur, H. Maxfield, J. Olmedo, A. Ori, F. Pretorius, F. Ramazanoglu, P. Singh and M. Varadarajan. I would also like to thank Tommaso De Lorenzo and Fethi Ramazanoglu for preparing most of the figures. This work was supported in part by the Eberly and Atherton funds of Penn State, USA and the  Distinguished Visiting Research Chair program of the Perimeter Institute, Canada.

\end{document}